\documentclass[11pt]{article}
\usepackage{amsmath,amsthm,amssymb,cite,enumerate}
\usepackage[a4paper,left=0.5in,right=1in]{geometry}
\usepackage[usenames]{color}
\usepackage{graphicx}
\usepackage{epsfig}
\usepackage{dcolumn}
\usepackage{bm}
\usepackage{authblk} 
\usepackage{soul} 
\usepackage{array,multirow} 

\usepackage{pifont} 

\begin{document}

\def \d {{\rm d}}

\def \bm #1 {\mbox{\boldmath{$m_{(#1)}$}}}

\def \bF {\mbox{\boldmath{$F$}}}
\def \bA {\mbox{\boldmath{$A$}}}
\def \bV {\mbox{\boldmath{$V$}}}
\def \bff {\mbox{\boldmath{$f$}}}
\def \bT {\mbox{\boldmath{$T$}}}
\def \bk {\mbox{\boldmath{$k$}}}
\def \bl {\mbox{\boldmath{$\ell$}}}
\def \bn {\mbox{\boldmath{$n$}}}
\def \bbm {\mbox{\boldmath{$m$}}}
\def \tbbm {\mbox{\boldmath{$\bar m$}}}
\def \l {\ell}

\def \bo {\mbox{\boldmath{$\omega$}}}
\def \bot {\mbox{\boldmath{$\tilde\omega$}}}
\def \bE {\mbox{\boldmath{$e$}}}
\def \bEt {\mbox{\boldmath{$\tilde e$}}}
\def \bG {\mbox{\boldmath{$\Gamma$}}}
\def \bGt {\mbox{\boldmath{$\tilde\Gamma$}}}

\def \T {\bigtriangleup}
\newcommand{\msub}[2]{m^{(#1)}_{#2}}
\newcommand{\msup}[2]{m_{(#1)}^{#2}}

\newcommand*\bR{\ensuremath{\boldsymbol{R}}}

\newcommand{\be}{\begin{equation}}
\newcommand{\ee}{\end{equation}}

\newcommand{\beq}{\begin{eqnarray}}
\newcommand{\eeq}{\end{eqnarray}}
\newcommand{\pa}{\partial}
\newcommand{\pp}{{\it pp\,}-}
\newcommand{\ba}{\begin{array}}
\newcommand{\ea}{\end{array}}

\newcommand{\M}[3] {{\stackrel{#1}{M}}_{{#2}{#3}}}
\newcommand{\m}[3] {{\stackrel{\hspace{.3cm}#1}{m}}_{\!{#2}{#3}}\,}

\newcommand{\tr}{\textcolor{red}}
\newcommand{\tb}{\textcolor{blue}}
\newcommand{\tg}{\textcolor{green}}

\newcommand*\bg{\ensuremath{\boldsymbol{g}}}
\newcommand*\bh{\ensuremath{\boldsymbol{h}}}

\def\a{\alpha}
\def\g{\gamma}
\def\de{\delta}
\def\b{\beta}

\def\E{{\cal E}}
\def\B{{\cal B}}
\def\R{{\cal R}}
\def\F{{\cal F}}
\def\L{{\cal L}}

\def\e{e}
\def\bb{b}

\newtheorem{theorem}{Theorem}[section] 
\newtheorem{cor}[theorem]{Corollary} 
\newtheorem{lemma}[theorem]{Lemma} 
\newtheorem{prop}[theorem]{Proposition}
\newtheorem{definition}[theorem]{Definition}
\newtheorem{remark}[theorem]{Remark}

\title{All non-expanding gravitational waves in $D$-dimensional (anti-)de~Sitter space}

\author[1]{Marcello Ortaggio\thanks{ortaggio(at)math(dot)cas(dot)cz}}
\author[2]{Jakub Vold\v{r}ich\thanks{kubavoldrich(at)seznam(dot)cz}}
\author[3,1]{Jos{\'e} Barrientos\thanks{jbarrientos(at)academicos(dot)uta(dot)cl}}

\affil[1]{Institute of Mathematics, Czech Academy of Sciences, \newline \v Zitn\' a 25, 115 67 Prague 1, Czech Republic}
\affil[2]{Institute of Theoretical Physics, Faculty of Mathematics and Physics, \newline
 Charles University, V Hole\v{s}ovi\v{c}k\'{a}ch 2, 180 00 Prague 8, Czech Republic}
\affil[3]{Sede Esmeralda, Universidad de Tarapac{\'a}, Avenida Luis Emilio Recabarren 2477, Iquique, Chile}

\maketitle

\begin{abstract}
 We present a complete, theory-independent classification of $D$-dimensional Kundt spacetimes of Weyl and traceless-Ricci type N. We show that these geometries consist of three invariantly defined subfamilies, namely (generalized) Kundt, {\it pp\,}- and Siklos waves, for each of which we obtain a convenient canonical form. As a byproduct, this also demonstrates that such metrics coincide with the class of non-expanding (A)dS-Kerr-Schild spacetimes. The role of these spacetimes in Einstein's gravity (including minimally coupled $p$-forms and non-linear electrodynamics) as non-expanding gravitational waves in an (anti)-de~Sitter background is discussed. Furthermore, applications to extended theories such as Gauss-Bonnet, Lovelock, quadratic and $f(R)$ gravity are also briefly illustrated, as well as the overlap of the obtained metrics with universal and almost-universal spacetimes. In the appendices we additionally settle the issue of the redundancy of certain field equations for all Kundt spacetimes in a theory-independent way, and present various alternative coordinates for the spacetimes studied in the paper.
\end{abstract}

\tableofcontents

\section{Introduction and summary of results}

\label{sec_intro}

Among exact vacuum solutions describing gravitational waves, the Kundt metrics of type~N have played a prominent role since the golden age of general relativity \cite{Kundt61,Kundt62}. They represent a generalization of plane waves also containing \pp waves \cite{Brinkmann25} as a special case, and are characterized by the presence of a congruence of twistfree, shearfree and non-expanding null geodesics  (cf. also the reviews \cite{Stephanibook,GriPodbook}). Further extensions to include a cosmological constant  were first systematically studied in \cite{GarPle81} (see also \cite{SalGarPle83,GarciaD83}). The complete family of Kundt Einstein spacetimes of type N was subsequently obtained and invariantly classified in \cite{OzsRobRoz85} (see also \cite{BicPod99I}). Type~N Kundt metrics can also be coupled with null Maxwell fields or with other types of light-like sources \cite{Peres60,Kundt61,Kundt62,Bonnor69,AicSex71,OzsRobRoz85,Siklos85,Stephanibook,GriPodbook},
 
Although originally obtained as solutions to Einstein's theory, various type N Kundt spacetimes were soon recognized to be also relevant to any $D$-dimensional theory which relies on a Lorenztian metric to describe gravity \cite{Deser75,Guven87,AmaKli89,HorSte90,Horowitz90,HorItz99}, also in the presence of certain modified electrodynamics \cite{Guven87,HorSte90,Horowitz90}. In fact, it has been proven more recently that all Einstein Kundt metrics of type N are ``universal'' \cite{HerPraPra14}, i.e., they simultaneously solve (virtually) any gravity theory with higher-order curvature corrections. Similarly, all Kundt spacetimes of (Weyl and traceless-Ricci) type~N can additionally accommodate null $p$-form fields which are immune to higher-order electromagnetic corrections \cite{OrtPra16,OrtPra18,HerOrtPra18} (some restrictions, however, emerge if backreaction is kept into account \cite{OrtPra16,KucOrt19}). They also have the peculiar property that all the scalar invariants constructed from the Riemann tensor and its covariant derivatives of arbitrary order are either vanishing or constant \cite{Pravdaetal02,Coleyetal04vsi,ColHerPel06} (cf. also footnote~4 of \cite{OrtPraPra10}). Furthermore, the past decade has seen a renewed interest in such spacetimes also in the context of the classical Kerr-Schild \cite{MonOCoWhi14,CarPenTro18,GurTek18,BahStaWhi20,OrtPraPra24} (cf. \cite{Kundt61,Siklos85} for related results in four dimensions) and Weyl \cite{Lunaetal19,Godazgaretal21,Han22} double copy.

The main purpose of the present paper is to achieve a complete classification of all $D$-dimensional {\em Kundt spacetimes of Weyl and traceless-Ricci type N} (as defined in \cite{Coleyetal04,Milsonetal05}, cf. also the reviews \cite{Coley08,OrtPraPra13rev}) in a theory-independent way. 
As it turns out, these geometries consist of three subfamilies characterized by the invariant sign of a ``kinematic'' quantity $\kappa$ (defined in eq.~\eqref{kappa} below). Additionally, after taking advantage of certain coordinate freedoms, we will achieve a canonical form for each of these subfamilies. Eventually, our analysis will reveal that under the assumptions described above, any metric in dimensions $D=2+n\ge 4$ can be written locally as
\be
 \d s^2=2\frac{Q^2}{P^2}\d u\left[\left(\frac{\kappa}{2}v^2+\frac{Q_{,u}}{Q}v\right)\d u+\d v\right]+P^{-2}\delta_{\a\b}\d x^\a\d x^\b-QP^{n/2-2}H\d u^2 \qquad (\a,\b=1,\ldots,n) ,
 \label{metric_final}
\ee
where  
\be
	 P=1+\frac{\lambda}{4}x_\a x^\a \quad (\lambda=\mbox{const}) ,
\label{P}
\ee	
and $H=H(u,x)$ is an arbitrary function (throughout the paper, $x$ stands collectively for $(x_1,\ldots,x_n)$). The form of the remaining function $Q$ and of the constant $\kappa$ are specified in table~\ref{tab_canonical} for the three possible invariant subfamilies of metrics.  The obtained line-element additionally reveals that all such metrics belong to the (A)dS-Kerr-Schild class (and in fact coincide with the family of non-expanding (A)dS-Kerr-Schild spacetimes, cf. section~\ref{subsec_summary}). The Ricci scalar is given by $R=D(D-1)\lambda$, and for $\lambda=0$ the three subfamilies contract to just two, i.e., the well-known type~N {\it pp\,}- and Kundt waves \cite{Schimming74,Coleyetal03,Coleyetal06,KucOrt19,OrtPraPra24}. In the special case $\kappa=0$, we will also relate our metrics to the Siklos waves discussed previously using a different coordinate system \cite{GibRub86,CveLuPop99,ChaGib00}.

\begin{table}[t]
	\[
	\begin{array}{|l|c|c|c|c|} \hline
		\mbox{Subfamily}  & \  \kappa \  & Q & \lambda & \mbox{comments}  \\ \hline\hline
		\mbox{g. Kundt waves}  & 1 & x_1 & \mbox{any} &  \\ \hline		
		\mbox{g. \pp waves} & \lambda & 1-\frac{\lambda}{4}x_\sigma x^\sigma & \mbox{any} &    \\  \hline		      
		\mbox{g. Siklos waves} & 0 & \frac{1}{4l^2}(x_\sigma+2l^2\b_\sigma)(x^\sigma+2l^2\b^\sigma)  & \lambda=-l^{-2}=-\beta_\sigma\beta^\sigma<0 & \b^\sigma=\b^\sigma(u)   \\ \hline      		  
	\end{array}
	\]
	\caption{\small~Invariant subfamilies of $D$-dimensional Kundt spacetimes of Weyl and traceless-Ricci type N (line-element~\eqref{metric_final}, \eqref{P}) and respective canonical forms of the metric functions. The abbreviation ``g.'' stands for ``generalized''. The constant $\lambda$ fixes the Ricci scalar $R=D(D-1)\lambda$. In the case $\kappa=0$, a special sublcass is obtained when $\b^\sigma=$const, which defines Siklos waves (a subcase of ``generalized'' Siklos waves; cf. also appendix~\ref{app_Siklos}). In the case $\lambda>0$, generalized Kundt waves and generalized \pp waves are locally isometric. The case $D=4$ corresponds to the results of \cite{OzsRobRoz85}.} 
	\label{tab_canonical}
\end{table}

It should be observed that, for $D=4$, from our conclusions one recovers the early results of \cite{OzsRobRoz85}. However, the extension of the latter to an arbitrary $D$ does not appear to be either straightforward nor a priori obvious (cf. a related comment after~\eqref{typeN_-1_trace} and, in particular, the analysis of section~\ref{sec_canonical}). Furthermore, it should also be mentioned that 
$D$-dimensional geometries of the form~\eqref{metric_final}, \eqref{P} have been already studied 
in \cite{Obukhov04,GleDot05}, in particular in the context of the Einstein-Maxwell and Lovelock-Yang-Mills theory, respectively. By analogy with the four-dimensional solutions of \cite{OzsRobRoz85}, references~\cite{Obukhov04,GleDot05} assumed metric~\eqref{metric_final}, \eqref{P} as an ansatz, but did not prove that it actually contains {\em all} $D$-dimensional Kundt spacetimes of Weyl and traceless-Ricci type N. This is instead one of the main results of the present contribution. Additionally, the emergence of Siklos waves as a special subcase of~\eqref{metric_final}, \eqref{P}, as mentioned above, was not discussed in \cite{Obukhov04,GleDot05}. More comments on the results of \cite{Obukhov04,GleDot05} will be given subsequently in section~\ref{sec_examples}.

The plan of the paper is as follows. In section~\ref{sec_kundt} we define the Kundt class of spacetimes, we explore the consequences of (some of) the curvature conditions implied by the type~N assumption, and introduce both a convenient normalization of the Kundt null vector field $\bl$ as well as suitable adapted coordinates (eq.~\eqref{metric}). In section~\ref{sec_N} the full type~N assumption is enforced, which fixes the metric functions $g_{\a\b}$, $F$ and $S$ (except for the $v$-independent part of $S$; eqs.~\eqref{g_canon}, \eqref{F}, \eqref{Q}, \eqref{S}, \eqref{S2}, \eqref{kappa}, \eqref{typeN_-1_S1}) and subjects $Z_\a$ to eq.~\eqref{typeN_-1_final}. In section~\ref{sec_canonical}, a remaining coordinate freedom is exploited to first arrive at canonical forms for the three sublclasses of metrics defined by the invariant sign of $\kappa$ (eq.~\eqref{kappa}), and then to show that in all cases $Z_\a$ can be transformed to zero while preserving such canonical forms. Finally, the role of such spacetimes as solutions of various gravity theories of particular interest is discussed in section~\ref{sec_examples}. Some additional technicalities are given in the appendices. Appendix~\ref{app_conserv} settles the issue of the redundancy of certain field equations for all Kundt spacetimes (i.e., not only those of Weyl and traceless-Ricci type N studied in the main body of the paper) in a theory-independent way, and is thus by itself of interest for possible further applications. Appendix~\ref{app_kundt_coords} relates the coordinates used in the paper for all Kundt spacetimes of Weyl and traceless-Ricci type N to the standard Kundt ones \cite{Kundt61,Kundt62,Stephanibook,GriPodbook}. Appendices~\ref{app_Siklos} and \ref{app_gen_pp} give the transformations 
to alternative coordinates often employed in the literature for certain subfamilies of Kundt metrics of Weyl and traceless-Ricci type N.

\section{Preliminaries: Kundt spacetimes}

\label{sec_kundt}

\subsection{Geometry of Kundt spacetimes and almost-Killing normalization}

A Kundt spacetime is characterized by admitting a congruence of null geodesics whose tangent vector field $\bl$ is twistfree, shearfree and expansionfree. This can be written concisely as
\be
 \l_{a;b}=\xi_a \l_b+\l_a\mu_b , \qquad \l_a\l^a=0=\xi_a\l^a .
 \label{kundt}
\ee
This definition is unaffected by an arbitrary rescaling $\bl\mapsto A\bl$ (it changes $\mu_a\mapsto\mu_a+(\ln A)_{,a}$, which just amounts to a redefinition of $\mu_a$). The contraction $\l_{c;a}\l^c_{\phantom{c};b}=(\xi^c\xi_c)\l_a\l_b$ defines a scalar quantity $\xi^c\xi_c$ which is invariant under such a rescaling and will be useful later on.

If an affine parametrization is chosen (always possible up to rescaling $\bl$), in addition to \eqref{kundt} one has also
\be
 \mu_a\l^a=0 ,
 \label{affine}
\ee
which will be assumed hereafter (constraining the remaining scaling freedom by $\l^aA_{,a}=0$). Using \eqref{kundt} and the Ricci identity one then finds
\be
 \frac{1}{2}\pounds_{\bl}(\pounds_{\bl}g_{ab})=(\xi^c\xi_c)\l_a\l_b+\l^c\l^dR_{dabc} .
 \label{LLg}
\ee

Computing $\l_{a;[bc]}$ from \eqref{kundt} and using the Ricci identity, it is also easy to see that 
\be
	\l_{[e}R_{a]b[cd}\l_{f]}\l^b=0 ,
	\label{I}
\ee	
which implies \cite{Ortaggio09} that the Riemann type of a Kundt spacetime is automatically I (or more special) \cite{ColHerPel06,OrtPraPra07,PodZof09} -- in particular the Weyl and the Ricci tensor share an aligned null direction.

In this work we are interested in a subset of Kundt spacetimes for which the Weyl and the traceless part of the Ricci tensor are both of type N, thus necessarily w.r.t. the same null direction $\bl$ \cite{OrtPraPra07}. This means that, in a null frame adapted to $\bl$, the only non-zero components of both these tensors are of boost weight $-2$ \cite{Coleyetal04,Milsonetal05,OrtPraPra13rev}. A necessary condition for this to happen is \cite{Ortaggio09}\footnote{More precisely, condition~\eqref{IId} defines the (aligned) Ricci type II and Weyl type II(d) \cite{Ortaggio09}, which in a null frame adapted to $\bl$ reads $R_{0i0j}=R_{0ijk}=R_{010i}=R_{01ij}=0$ \cite{Coleyetal04}. Condition~\eqref{I} instead only means $R_{0i0j}=R_{0ijk}=0$ \cite{Ortaggio09}.\label{footn_typeII}}
\be
	R_{ab[cd}\l_{e]}\l^b=0 ,
	\label{IId}
\ee
which also implies 
\be
 \l^c\l^dR_{dabc}={\cal R} \l_a\l_b ,
 \label{Rll}
\ee
where ${\cal R}$ is at this stage an unspecified function (invariant under a rescaling of $\bl$). From $2\xi^a\l^b\l_{a;[bc]}=-\xi^a\l^b\xi_{a;b}\l_c$, using the Ricci identity and \eqref{Rll} one concludes
\be
 \pounds_{\bl}(\xi^c\xi_c)=0 .
 \label{Luu}
\ee

The freedom of rescaling $\bl$ is often employed to rewrite~\eqref{kundt} in the ``canonical'' form $\l_{a;b}=\xi_a \l_b+\l_a\xi_b$, around which standard Kundt coordinates are naturally constructed \cite{PodOrt06,PodZof09}. However, when \eqref{IId} is satisfied, a different normalization turns out to be more convenient, as observed in \cite{OzsRobRoz85} in four dimensions. Let us show this. 

First, the identity $\l_{[a,bc]}=0$ with~\eqref{kundt} gives
\be
 \l_{[a}(\xi_{b,c]}-\mu_{b,c]})=0 ,
 \label{c1}
\ee 
while the Ricci identity reveals that \eqref{IId} holds iff
\be
 \l_{[a}\mu_{b,c]}=0 .
 \label{c2}
\ee 

Combining \eqref{c1} and \eqref{c2} gives $\l_{[a}(\xi_{b,c]}+\mu_{b,c]})=0$, so that by Frobenius' theorem one can write
\be
 \xi_a+\mu_a=L\l_a+f_{,a} ,
\ee
where $L$ and $f$ are scalar functions with $\l^af_{,a}=0$. This means that $\l_{a;b}=L\l_a\l_b+\l_af_{,b}+\xi_a\l_b-\l_a\xi_b$, where the term $\l_af_{,b}$ 
can be removed by a rescaling of $\bl$ (cf. the comment after~\eqref{kundt}). One thus finally arrives at
\be
 \l_{a;b}=L\l_a\l_b+\xi_a\l_b-\l_a\xi_b .
 \label{almostK}
\ee

This is an ``almost-Killing'' normalization \cite{OzsRobRoz85}, in the sense that 
\be
 \pounds_{\bl}g_{ab}=2L\l_a\l_b ,
 \label{Lg}
\ee
so that $\bl$ is Killing iff $L=0$. 
The Lie derivative of \eqref{Lg} gives
 \be
 \frac{1}{2}\pounds_{\bl}(\pounds_{\bl}g_{ab})=(\pounds_{\bl}L)\l_a\l_b ,
\ee
hence by \eqref{LLg}, \eqref{Rll} one gets
\be
 \pounds_{\bl}L=(\xi^c\xi_c)+{\cal R} . 
\ee

There is still a remaining freedom $\bl\mapsto A\bl$ with $\d A\propto\bl$. This leaves \eqref{almostK} unchanged, provided $L$ is suitably redefined.  However, the scalar quantity
\be
 \hat L\equiv \pounds_{\bl}L ,
\ee
is an invariant whose sign will prove useful in classifying distinct families of solutions (in section~\ref{sec_canonical}, with the definition~\eqref{kappa}). Recalling~\eqref{Luu} one has $\pounds_{\bl}\hat L=\pounds_{\bl}{\cal R}$.

\subsection{Adapted coordinates}

Similarly as for the canonical Kundt line-element constructed in \cite{PodOrt06,PodZof09}, one can now introduce adapted coordinates $(u,v,x)$ (where $x$ stands collectively for $(x_1,\ldots,x_n)$), where $v$ is an affine parameter along $\bl$, such that $\bl=\pa_v$, and $u$=const defines null hypersurfaces to which $\bl$ is orthogonal. However, the normalization \eqref{almostK} implies that here generically the covector $\l_a\d x^a$ will not be a gradient (as opposed to \cite{PodOrt06,PodZof09}), i.e., $\l_a\d x^a=F\d u$, where $F$ is a spacetime function. Apart from this difference, one can follow the same steps as in \cite{PodOrt06,PodZof09} to obtain the line-element
\be
 \d s^2=2F\d u(S\d u+\d v+Z_\a\d x^\a)+g_{\a\b}\d x^\a\d x^\b ,
 \label{metric}
\ee
where $\a,\b=1,\ldots,n$, while \eqref{Lg} implies 
\be
 F_{,v}=0 , \qquad Z_{\a,v}=0 , \qquad g_{\a\b,v}=0 , \qquad S_{,v}=LF ,
\ee
along with 
\be
 \hat L=F^{-1}S_{,vv} .
 \label{Svv_L}
\ee

The above metric represents the most general Kundt spacetime of Weyl type II(d) and traceless-Ricci type II. Below we will obtain further restrictions which reduce  both tensor to the more special type N.

\section{Type N condition}

\label{sec_N}

In order to specialize the line-element \eqref{metric} to the algebraic type~N we need to compute the corresponding curvature tensor. Let us introduce the following adapted null co-frame
\be
 \bo^0=S\d u+\d v+Z_\a\d x^\a , \qquad \bo^1=F\d u , \qquad \bo^i=\bot^i \quad (i,j\ldots=2,\ldots,D-1) ,
\ee
where, from now on, a tilde refers to quantities intrinsic to the transverse metric $g_{\a\b}$. The dual frame reads
\be
 \bE_0=\pa_v , \qquad \bE_1=F^{-1}(\pa_u-S\pa_v) , \qquad \bE_i=\bEt_i-Z_i\pa_v ,
 \label{coframe}
\ee   
where $\bot^i=\tilde\omega^i_{\a}\d x^\a$, $\bEt_i=\tilde e^\a_i\pa_{x^\a}$ and $Z_i=Z_\a\tilde e^\a_i$ (note that $\bE_0=\bl$).

Straightforward computations produce the connection 1-forms\footnote{We follow the notation of \cite{Stephanibook}. Recall $\bG^0_{\phantom{0}0}=-\bG^1_{\phantom{1}1}$, $\bG^i_{\phantom{i}0}=-\bG^1_{\phantom{1}i}$, $\bG^i_{\phantom{i}1}=-\bG^0_{\phantom{0}i}$.}
\beq
 & & \bG^1_{\phantom{1}1}=-\frac{1}{2}F^{-1}F_{|i}\bot^i-F^{-1}S_{,v}\bo^1 , \qquad \bG^1_{\phantom{1}i}=\frac{1}{2}F^{-1}F_{|i}\bo^1 , \label{G11} \\
 & & \bG^0_{\phantom{0}i}=\frac{1}{2}F^{-1}F_{|i}\bo^0+F^{-1}(S_{|i}-S_{,v}Z_i-m_i)\bo^1-(Z_{[i||j]}+F^{-1}n_{(ij)})\bo^j , \\
 & & \bG^i_{\phantom{i}j}=(Z_{[i||j]}-F^{-1}n_{[ij]})\bo^1+\bGt^i_{\phantom{i}j} , \label{Gij} 
\eeq
where $F_{|i}=e^\a_iF_{,\a}$, a double bar denotes covariant differentiation in the transverse space, and we have defined
\be
 m_i=Z_{\a,u}\tilde e^\a_i , \qquad n_{ij}=\tilde \omega^i_{\a,u} \tilde e^\a_j .
 \label{n_m}
\ee
(Note that $2n_{(ij)}=\tilde e^\a_i\tilde e^\b_j g_{\a\b,u}$.)

Using \eqref{G11}--\eqref{Gij} one can then compute the non-zero components of the Riemann tensor (cf. footnote~\ref{footn_typeII}; note that now ${\cal R}=-R_{0101}$)
\beq
 & & R_{0101}=\frac{1}{4}\left(\ln F\right)_{|i}\left(\ln F\right)_{|i}-F^{-1}S_{,vv} , \ \ R_{0i1j}=-\frac{1}{2}\left(\ln F\right)_{||ij}-\frac{1}{4}\left(\ln F\right)_{|i}\left(\ln F\right)_{|j} , \ \ \, R_{ijkl}=\tilde R_{ijkl} , \label{Riem0} \\
 & & R_{011i}=F^{-1}\left[S_{|i,v}-S_{,vv}Z_i-\frac{1}{2}\left(\ln F\right)_{|i,u}+\frac{1}{2}\left(\ln F\right)_{|j}\left(n_{[ij]}-FZ_{[i||j]}\right) \right] , \label{Riem-1i}\\
 & & R_{1ijk}=F^{-3/2}\left[-\big(F^{3/2}Z_{[j||k]}\big)_{||i}+\big(F^{3/2}\big)_{|[i}Z_{j||k]}\right]+F^{-1/2}\left[\big(F^{-1/2}n_{(ij)}\big)_{||k}-\big(F^{-1/2}n_{(ik)}\big)_{||j}\right] , \label{Riem-1ijk}  \\
 & & R_{1i1j}=F^{-1}\bigg[F^{-1}F_{|(i}\left(Z_{j)}S_{,v}-S_{|j)}+m_{j)}\right)-S_{,vv}Z_iZ_j+S_{,v}\left(Z_{(i||j)}-F^{-1}n_{(ij)}\right)-S_{||ij}+2Z_{(i}S_{|j),v}  \nonumber \\
 & & \qquad\qquad\qquad  {}+m_{(i||j)}-\frac{1}{2}\tilde e^\a_i\tilde e^\b_j\left(F^{-1}g_{\a\b,u}\right)_{,u}\bigg]+\left(Z_{[i||k]}-F^{-1}n_{(ik)}\right)\left(Z_{[j||k]}-F^{-1}n_{(jk)}\right) . \label{Riem-2}
\eeq

Now, in the null frame defined above the Weyl and traceless-Ricci type N condition reads simply 
\beq
 & & R_{0101}=-\lambda , \qquad R_{0i1j}=\lambda\delta_{ij} , \qquad R_{ijkl}=\lambda(\delta_{ik}\delta_{jl}-\delta_{il}\delta_{jk}) , \label{typeN_0} \\
 & & R_{011i}=0 , \qquad R_{1ijk}=0 \label{typeN_-1} ,
\eeq
where 
\be
 \lambda\equiv\frac{R}{D(D-1)} 
 \label{lambda}
\ee
is necessarily a constant (as follows plugging \eqref{typeN_0}, \eqref{typeN_-1} into the the Bianchi identities (B.1), (B.12) and (B.13) of \cite{Pravdaetal04}; cf. also the proof of Proposition~A.8 of \cite{HerOrtPra18}).\footnote{In particular, in general relativity with a cosmological constant $\Lambda$ one has $\lambda=\frac{2\Lambda}{(D-1)(D-2)}$.}

The first of \eqref{typeN_0} with \eqref{Riem0} gives
\be
 S=S_2v^2+S_1v+S_0 , 
 \label{S}
\ee
with
\be
 S_2=\frac{1}{2}F\left[\lambda+\frac{1}{4}\left(\ln F\right)_{|i}\left(\ln F\right)_{|i}\right] , \qquad S_{1,v}=0=S_{0,v} .
 \label{S2}
\ee
With \eqref{Svv_L} this also shows
\be
 \hat L=\lambda+\frac{1}{4}\left(\ln F\right)_{|i}\left(\ln F\right)_{|i} ,
\ee
and $\pounds_{\bl}\hat L=0$.

The third of \eqref{typeN_0} with \eqref{Riem0} means that the spatial metric $g_{\a\b}$ must be of constant curvature, with $\tilde R=n(n-1)\lambda$. Without losing generality, a coordinate transformation of the form $x_\a\mapsto x_\a'(u,x)$ can be used to cast $g_{\a\b}$ in the canonical form \cite{Stephanibook}
\be
 g_{\a\b}=P^{-2}\delta_{\a\b} , \qquad P=1+\frac{\lambda}{4}x_\a x^\a , 
 \label{g_canon}
\ee
while keeping the line-element~\eqref{metric} otherwise unchanged. From now on, Greek indices $\a,\b,\ldots$ will be lowered and raised with the flat metric $\delta_{\a\b}$ or $\delta^{\a\b}$ (so that, e.g., $x_\a x^\a=x_1^2+\ldots+x_n^2$).

The form~\eqref{g_canon} results in significant simplifications, for the spatial connection 1-forms now reduce to
\be
 \bGt^i_{\phantom{i}j}=-\left(\ln P\right)_{|j}\bo^i+\left(\ln P\right)_{|i}\bo^j ,
 \label{Gij_simplified}
\ee
so that $Z_{i||j}=Z_{i|j}+Z_j\left(\ln P\right)_{|i}-Z_k\left(\ln P\right)_{|k}\delta_{ij}$, and
\be
 n_{ij}=0 .
 \label{n}
\ee

In order to solve the second of \eqref{typeN_0}, it is convenient to define an auxiliary function $Q(u,x)$ such that
\be
 F=\frac{Q^2}{P^2} . 
\label{F}
\ee
Using this and the second of \eqref{Riem0}, it is not difficult to arrive at the solution
\be
 Q=\left(1-\frac{\lambda}{4}x_\sigma x^\sigma\right)\a+\b_\sigma x^\sigma ,
 \label{Q}
\ee
where $\a=\a(u)$ and $\b_\sigma=\b_\sigma(u)$ (the line-element is invariant under $Q\mapsto-Q$, therefore we hereafter assume $\a\ge0$ without losing generality).\footnote{For later purposes, let us observe that (as a consequence of~\eqref{typeN_0}, \eqref{Riem0}, \eqref{F}, \eqref{g_canon} and \eqref{coframe}) $Q$ satisfies the condition $\left[\left(\ln Q\right)_{||ij}+\left(\ln Q\right)_{|i}\left(\ln Q\right)_{|j}-2\left(\ln P\right)_{|(i}\left(\ln Q\right)_{|j)}\right]_{,u}=0$.\label{foot_Q}} With \eqref{g_canon}, this implies, in particular, that necessarily $F_{|i}\neq0$ when $\lambda\neq0$. Furthermore, using \eqref{g_canon}, \eqref{F} and \eqref{Q}, eq.~\eqref{S2} becomes
\be
 \kappa\equiv F\hat L=2S_2=\lambda\a^2+\b_\sigma \b^\sigma .
 \label{kappa}
\ee

Next, the first of \eqref{typeN_-1} with \eqref{Riem-1i} and \eqref{S} gives 	
\be
S_{1|i}-2S_2Z_i-\left(\ln Q\right)_{|i,u}-\frac{1}{2}\left(\ln F\right)_{|j}FZ_{[i||j]}=0 .
 \label{typeN_-1_S1}
\ee

Finally, the last condition to impose is the second of \eqref{typeN_-1} with \eqref{Riem-1ijk} and \eqref{n}, giving 
\be
 \big(F^{3/2}Z_{[j||k]}\big)_{||i}=\big(F^{3/2}\big)_{|[i}Z_{j||k]} .
 \label{typeN_-1_final}
\ee
The integration of this equation will be discussed in section~\ref{sec_canonical}, taking advantage of certain canonical forms of the metric function $Q$ which will be obtained there.

Tracing over $k$ and $i$ one arrives at 
\be
 \left(F^{3/2}Z_{[j||k]}\right)^{||k}=0 .
 \label{typeN_-1_trace}
\ee

When $n=2$, the tensorial eqs. \eqref{typeN_-1_final} and \eqref{typeN_-1_trace} are equivalent (each reduces to just two ``scalar'' equations) and mean that the 2-form $F^{3/2}Z_{[j||k]}$ must be covariantly constant in the 2-geometry of $g_{\a\b}$, hence just constant, since $n=2$ (in coordinate components, this means that $Q^3P^{-1}Z_{[\a,\b]}$ can only depend on $u$). However, for $n>2$ this is no longer the case, therefore the argument used in \cite{OzsRobRoz85} to set $Z_\a=0$ does not apply in higher dimensions. Nevertheless, we will show in section~\ref{sec_canonical} how one can arrive at the same conclusion in arbitrary dimension.

\section{Invariant subfamilies and their canonical forms}

\label{sec_canonical}

\subsection{Coordinate freedom}

\subsubsection{Redefinitions of $u$ and $v$}

First, let us note that under a transformation (cf. also \cite{Obukhov04})
\be
 u=u(u')  \qquad (\d u/\d u'>0) ,
 \label{u_redef}
\ee
one has 
\be
 Q'^2=Q^2\frac{\d u}{\d u'}  , \qquad S'= S \frac{\d u}{\d u'} .
\ee
This freedom will be useful in the following to normalize $\a$ or one of the $\b_\sigma$ in~\eqref{Q}.

One can further redefine the affine parameter $v$ as
\be
 v=h(u)v'+f(u,x) \qquad (h>0) , 
 \label{v_redef}
\ee
resulting in (a dot denoting differentiation w.r.t. $u$)
\beq
 & & P'=P , \qquad Q'^2=hQ^2 , \qquad Z'_\a=h^{-1}(Z_\a+f_{,\a}) , \nonumber \\
 & & S_2'=hS_2 , \qquad S_1'=S_1+h^{-1}\dot h+2fS_2 , \qquad S_0'=h^{-1}(S_0+\dot f+f^2S_2+fS_1) . \label{v_redef_Z}
\eeq
This shows, for example, that if $S_1=S_1(u)$, then it can be removed, e.g., by a transformation with $f=0$.

\subsubsection{Rotations}

Furthermore, the transverse space metric~\eqref{g_canon} is invariant under $u$-dependent spatial rotations
\be
 x_\a=M_\a^{\phantom{\a}\b}(u) x'_\b \qquad (M_\a^{\phantom{\a}\b}M^\a_{\phantom{\a}\g}=\delta^\b_\g) ,
 \label{rotation}
\ee
which gives rise to a spacetime line-element of the same form as the original one but with $P'=1+\frac{\lambda}{4}x'_\sigma x'^\sigma$, $Q'=\left(1-\frac{\lambda}{4}x'_\sigma x'^\sigma\right)\a'+\b'_\sigma x'^\sigma$ and
\beq
 & & \a'=\a , \qquad \b'_\g=M^\sigma_{\phantom{\sigma}\g}\b_\sigma , \qquad Z'_\a=M^\b_{\phantom{\b}\a}(Z_\b+Q^{-2}\dot M_\b^{\phantom{\b}\g}M^\sigma_{\phantom{\sigma}\g}x_{\sigma}) , \nonumber \\
 & & S_0'=S_0+\dot M^\a_{\phantom{\a}\b} M_\sigma^{\phantom{\sigma}\b}x^\sigma\left(Z_\a+\frac{1}{2}Q^{-2}\dot M_\a^{\phantom{\a}\g}M^\rho_{\phantom{\rho}\g}x_\rho\right) ,
\eeq
with the remaining metric functions unchanged.
For later purposes, let us observe that the 2-form $Z_{[\alpha,\rho]}$ thus transforms as
\be
     M_{\beta}^{\phantom{\beta}\alpha} M_{\tau}^{\phantom{\tau}\rho}\pa'_{[\rho}Z'_{\alpha]} = \pa_{[\tau}Z_{\beta]} + (Q^{-2}x_\sigma)_{,[\tau} \Dot{M}_{\beta]}^{\phantom{a}\gamma}M^\sigma_{\phantom{a}\gamma}  , 
		\label{dZ_rot}
\ee
where, to avoid ambiguities, we have indicated explicitly the derivative operators $\pa'_{\rho}=\pa/\pa x'^\rho$ and $\pa_{\tau}=\pa/\pa x^\tau$.

\subsubsection{M\"obius transformations}

Finally, the spatial metric~\eqref{g_canon} is also invariant under the (subset of) M\"obius transformations (cf., e.g., \cite{Ahlfors81})\footnote{In the case $\lambda=0$ this has to be replaced by a translation $x_\sigma=a_\sigma+x_\sigma'$.\label{foot_transl}}
\beq
 & & x_\sigma=a_\sigma+\left(x_\sigma'-a_\sigma\right)\frac{\Delta}{\Sigma^2} , \label{Moebius} \\
 & & \Sigma^2\equiv(x'_{\rho}-a_{\rho})(x'^{\rho}-a^{\rho}) =\frac{\Delta^2}{(x_{\rho}-a_{\rho})(x^{\rho}-a^{\rho})} , \qquad \Delta\equiv\frac{4+\lambda a_\rho a^\rho}{\lambda} , \nonumber
\eeq
where $a_\sigma=a_\sigma(u)$ (and such that $\Delta\neq0$), resulting in a spacetime line-element with 
\beq
 & & \a'=\frac{\a\left(\lambda a_\rho a^\rho-4\right)-4\b_\rho a^\rho}{\lambda\Delta} , \qquad \b'_\sigma=\b_\sigma-2\frac{2\a+\b_\rho a^\rho}{\Delta}a_\sigma , \nonumber \\
 & & Z'_\a=\frac{\Delta}{\Sigma^2}\left[Z_\a-(x'_\a-a_\a)\frac{2Z_\b(x'^\b-a^\b)}{\Sigma^2}\right]+\frac{1}{\Delta Q'^2}\left[-2(x'_\a-a_\a)x'_\sigma\dot a^\sigma+\dot a_\a(\Sigma^2-\Delta)\right ] , \label{Moebius_Z'} \\
 & & S_0'=S_0+\frac{Z_\a}{\Sigma^4}\left[2(x'^\a-a^\a)\Psi+\dot a^\a\Sigma^2(\Sigma^2-\Delta)\right]+\frac{\dot a_\a}{2Q'^2\Delta^2}\left[4x'^\a\Psi+\dot a^\a(\Sigma^2-\Delta)^2\right] , \nonumber
\eeq
where we have defined
\be
 \Psi\equiv a_\sigma\dot a^\sigma(\Sigma^2-\Delta)+x'_\sigma\dot a^\sigma\Delta =\Sigma^2\dot a^\sigma x_\sigma.
\ee

(In order to check the above formulae it is useful to observe that here $P'\neq P$ and $Q'\neq Q$, however $Q'/P'=Q/P$, and $P'=P\Sigma^2/\Delta$.) In this case, the transformed 2-form $Z_{[\alpha,\rho]}$ can be written as
\beq
     Q'^3\pa'_{[\rho}Z'_{\alpha]} = & & \frac{\Sigma^2}{\Delta}Q^3\pa_{[\rho}Z_{\alpha]}+2\frac{\Sigma^4}{\Delta^3}(x^\sigma-a^\sigma)Q^3\left[(x_\a-a_\a)\pa_{[\sigma}Z_{\rho]}-(x_\rho-a_\rho)\pa_{[\sigma}Z_{\a]}\right] \nonumber \\
		& & {}+4\frac{\Sigma^2}{\Delta^2}(x_{[\a}-a_{[\a}) \left\{\dot a^\sigma\left[a_\sigma+\frac{\Sigma^2}{\Delta} (x_\sigma-a_\sigma)\right]\pa_{\rho]}Q-\dot a_{\rho]}\left[\frac{\Sigma^2-\Delta}{\Delta}(x^\sigma-a^\sigma)\pa_\sigma Q+Q\right]\right\} \nonumber \\
		& & {}-2\frac{\Sigma^2-\Delta}{\Delta} \dot a_{[\a}\pa_{\rho]}Q .
		\label{dZ_Moeb}
\eeq

Note that under \eqref{rotation} and \eqref{Moebius} one has $\kappa'=\kappa$, while $\kappa$ gets rescaled by a positive factor under both \eqref{u_redef} and \eqref{v_redef}. Depending on the (invariant) signs of $\kappa=\lambda\a^2+\b_\sigma \b^\sigma$ and $\lambda$, the above freedoms can be used to achieve the canonical forms summarized in table~\ref{tab_canonical}, as detailed in what follows.

\subsection{$\kappa>0$: (generalized) Kundt waves}

\label{subsec_k>0}

For $\kappa>0$ there are three possible cases to be treated separately, however, they all lead to the canonical form
\be
 \kappa=1 , \qquad Q=x_1 ,
 \label{Q1}
\ee
as we now explain.

First, if $\lambda\a\neq0$ a transformation~\eqref{Moebius} with\footnote{The sign of the quantity $\pm\sqrt{\kappa}$ can be chosen arbitrarily. This freedom is useful in order to prevent the transformation from becoming singular in the special cases $\beta_1=\mp\sqrt{\kappa}$. The final sign of $\beta'_1$ can be chosen as desired using $x'_1\mapsto-x'_1$.} 
\be
 a_1=2\frac{\beta_1\pm\sqrt{\kappa}}{\lambda\a} , \qquad a_{\hat\sigma}=\frac{2}{\lambda\a}\beta_{\hat\sigma} \quad (\hat\sigma\neq1) , 
 \label{1_generic}
\ee
leads to $\a'=0$, $\b'_\sigma=\mp\sqrt{\kappa}\delta_{\sigma,1}$. A further redefinition of $u$~\eqref{u_redef} can be used to arrive at (hereafter dropping the primes over the transformed quantities) $\beta_1=1$, so that finally~\eqref{Q1} is indeed achieved.

If $\a=0$, one can again arrive at~\eqref{Q1} using a spatial rotation~\eqref{rotation} followed by \eqref{u_redef}.

Finally, if $\lambda=0\neq\a$ one can use a translation $x_1=x_1'-\a/\beta_1$ (recall footnote~\ref{foot_transl}) followed by a rotation~\eqref{rotation} and \eqref{u_redef} to again obtain~\eqref{Q1}.

Next, thanks to \eqref{Q1}, the remaining type~N condition~\eqref{typeN_-1_final} can be integrated more easily. Recalling \eqref{coframe}, \eqref{g_canon}, \eqref{Gij_simplified}, \eqref{F} and naturally aligning the spatial frame vectors $\bEt_i$ along the coordinate vectors $\pa/\pa x^\a$, after some straightforward calculations one obtains (we refer to \cite{Voldrich_master} for details)
\be
 Z_{[\hat\a,\hat\b]}=\frac{2\lambda b_{[\hat\b}x_{\hat\a]}+c_{\hat\a\hat\b}}{(x_1)^2} , \qquad Z_{[1,\hat\a]}=\frac{2P}{(x_1)^3}b_{\hat\a}-\frac{1}{x_1}x_{\hat\sigma}Z_{[\hat\sigma,\hat\a]} ,
 \label{dZ_Kundt}
\ee
where coordinate indices with a hat are greater than 1, and $b_{\hat\a}$, $c_{\hat\a\hat\b}=-c_{\hat\b\hat\a}$ are integration functions depending only on $u$.

Note that the canonical form~\eqref{Q1} is invariant under a subset of spatial rotations~\eqref{rotation} specified by
\be
 M^1_{\phantom{1}\g}=\delta^1_\g , 
\label{rot_Kundt}
\ee
under which the transformed 2-form~\eqref{dZ_rot} takes the same form as the original one~\eqref{dZ_Kundt}, but with $b_{\hat\b}$ and $c_{\hat\a\hat\b}$ replaced by
\beq
  & & b'_{\hat\b}=M^{\hat\a}_{\phantom{\a}\hat\b} b_{\hat\a} , \label{b'_rot} \\
	& & c'_{\hat\a\hat\b}=M^{\hat\g}_{\phantom{\g}\hat\a}\left(M_{\hat\sigma}^{\phantom{\sigma}\hat\b}c_{\hat\g\hat\sigma}+\dot M_{\hat\g}^{\phantom{\g}\hat\b}\right) . \label{c'_rot}
\eeq

Additionally, eq.~\eqref{Q1} is also invariant under M\"obius transformations~\eqref{Moebius} with
\be
 a_1=0 .
\label{Moeb_Kundt}
\ee
In this case, the transformed 2-form~\eqref{dZ_Moeb} reads as in~\eqref{dZ_Kundt}, but now with $b_{\hat\b}$ and $c_{\hat\a\hat\b}$ replaced by\footnote{Here we assume $\lambda\neq0$ - cf. \cite{OrtPraPra24} for the case $\lambda=0$.}
\beq
  & & b'_{\hat\b}=-\frac{2}{\Delta}\left[b_{\hat\b}\left(\frac{2}{\lambda}-\frac{1}{2}a_{\hat\sigma}a^{\hat\sigma}\right)+a_{\hat\b}a_{\hat\sigma}b^{\hat\sigma}+\frac{1}{\lambda}\left(c_{\hat\b\hat\sigma}a^{\hat\sigma}+\dot a_{\hat\b}\right)\right] , \label{b'_Moeb} \\
	& & c'_{\hat\a\hat\b}=c_{\hat\a\hat\b}+\frac{4}{\Delta}a_{[\hat\a}\left(4b_{\hat\b]}+c_{\hat\b]\hat\sigma}a^{\hat\sigma}+\dot a_{\hat\b]}\right) .
\eeq

Combining the above freedoms, we can perform a coordinate transformation depending on $n(n-1)/2$ arbitrary functions of $u$ (i.e., $M^{\hat\a}_{\phantom{\a}\hat\b}$ and $a_{\hat\sigma}$), which is precisely the number of integration functions ($b_{\hat\b}$ and $c_{\hat\a\hat\b}$) appearing in~\eqref{dZ_Kundt}. We first use~\eqref{Moebius}, \eqref{Moeb_Kundt} to set $b'_{\hat\b}=0$ in~\eqref{b'_Moeb}, which means solving a system of first order, quasi-linear ODEs for the unknowns $a_{\hat\b}(u)$. If one chooses an initial condition for $a_{\hat\b}(u)$, Peano's theorem (cf., e.g., \cite{PicStaVidbook}) guarantees the local existence of a solution provided $b_{\hat\b}(u)$ and $c_{\hat\a\hat\b}(u)$  nuous functions (the solution does not need to be unique, but this is irrelevant for our purposes). Next, dropping the primes on the thus obtained transformed quantities, we can further use~\eqref{rotation}, \eqref{rot_Kundt} to set $c'_{\hat\a\hat\b}=0$ in~\eqref{c'_rot} by solving a system of first order, linear, homogeneous ODEs for the unknowns $M_{\hat\g}^{\phantom{\g}\hat\b}$, to which standard existence results apply \cite{PicStaVidbook,Croninbook}.\footnote{Namely, the system of ODEs is given by $\Dot{M}_{\hat\a}^{\phantom{a}\hat\rho}=-c_{\hat\a\hat\sigma}M_{\hat\sigma}^{\phantom{\sigma}\hat\rho}$, from which simple manipulations give $(M_{\hat\a}^{\phantom{\a}\hat\b}M^{\hat\a}_{\phantom{\a}\hat\g})\dot{}=0$ -- this guarantees that if one chooses initial conditions (e.g., at $u=0$) such that $M_{\hat\a}^{\phantom{\a}\hat\b}(0)$ is an orthogonal matrix, the corresponding solution $M_{\hat\a}^{\phantom{\a}\hat\b}(u)$ will automatically be orthogonal for any value of $u$.\label{footn_eq_M}}

Eventually (dropping again the primes on the transformed quantities), we have thus arrived at $\pa_{[\rho}Z_{\alpha]}=0$. This means that $Z_\a$ is a gradient, which ensures we can use a further transformation~\eqref{v_redef}, \eqref{v_redef_Z} (with $h=1$) to arrive at
\be
 Z'_\a=0 .
\ee

The metrics of this branch reduce to the so called Kundt waves when $\lambda=0$ (cf. \cite{Coleyetal03,Coleyetal06,OrtPraPra24} and appendix~\ref{app_kundt_coords}), while for $\lambda\neq 0$ they can be referred to as ``generalized Kundt waves'' (following the nomenclature used in four dimensions in \cite{GriDocPod04}). For $D=4$, these families were first obtained, respectively, in \cite{Kundt61} and \cite{GarPle81}, cf. also \cite{OzsRobRoz85,BicPod99I,Stephanibook,GriDocPod04,GriPodbook}.

Let us observe that for $\lambda>0$ a different, but equivalent, canonical form is also possible, as discussed in section~\ref{subsec_genpp} below.

\subsection{$\lambda\kappa>0$: generalized \pp waves}

\label{subsec_genpp}

When $\lambda\kappa>0$ one can always arrive at the canonical form
\be
 \kappa=\lambda , \qquad Q=1-\frac{\lambda}{4}x_\sigma x^\sigma ,
 \label{Q2}
\ee
as we now show in two different possible cases.

First, if $\b_\sigma\neq0$, a transformation~\eqref{Moebius} with 
\be
 a_{\sigma}=2\frac{-\lambda\a\pm\sqrt{\lambda\kappa}}{\lambda\beta_\rho\beta^\rho}\beta_{\sigma}  , 
\ee
gives $\a'=\mp\sqrt{\lambda\kappa}/\lambda$, $\b'_\sigma=0$. Next, using~\eqref{u_redef} one arrives at (again dropping the primes) $\a=1$, i.e., at \eqref{Q2}.

If, instead, $\b_\sigma=0$, it suffices to use~\eqref{u_redef} to normalize $\a=1$ and thus obtain again~\eqref{Q2}.

Next, using~\eqref{Q2} one can integrate~\eqref{typeN_-1_final} to obtain \cite{Voldrich_master}
\be
  Z_{[\a,\b]}=(Q^{-2}x_\sigma)_{,[\b}c_{\a]\sigma}  , 
	\label{dZ_pp}
\ee
where $c_{\a\sigma}=-c_{\sigma\a}$ are integration functions depending only on $u$.

Thanks to this result, we can now use a rotation~\eqref{rotation} to set $\pa'_{[\rho}Z'_{\alpha]}=0$. Namely, plugging~\eqref{dZ_pp} into~\eqref{dZ_rot} shows that $\pa'_{[\rho}Z'_{\alpha]}=0$ is equivalent to choosing the rotation matrix $M^\sigma_{\phantom{a}\gamma}$ such that $c_{\a\sigma}+\Dot{M}_{\a}^{\phantom{a}\gamma}M^\sigma_{\phantom{\sigma}\gamma}=0$. Using the orthogonality of $M^\sigma_{\phantom{a}\gamma}$, this can be rewritten as $ 
 \Dot{M}_{\a}^{\phantom{a}\rho}=-c_{\a\sigma}M_\sigma^{\phantom{\sigma}\rho}$, which is (essentially) the same system of ODEs as discussed in footnote~\ref{footn_eq_M}, therefore no further comments are required here.

From now on we drop the primes over the quantities transformed under rotations~\eqref{rotation}. Having achieved above $\pa_{[\rho}Z_{\alpha]}=0$, as in section~\ref{subsec_k>0}
 we can further use~\eqref{v_redef}, \eqref{v_redef_Z} to arrive at $Z'_\a=0$.

The metrics of this branch  can be  referred to as ``generalized \pp waves'', since they reduce to \pp waves (cf. section~\ref{subsubsec_pp}) in the limit $\lambda\to0$. The canonical form~\eqref{Q2} is particularly relevant to the case $\kappa<0$ with $\lambda<0$, since the alternative form~\eqref{Q1} is also possible in the case $\kappa>0$ with $\lambda>0$. First results in four dimensions date back to \cite{OzsRobRoz85} (cf. also \cite{BicPod99I,GriDocPod04}).

\subsection{$\kappa=0$: (generalized) Siklos waves and \pp waves}

\label{subsec_Siklos}

The case $\kappa=0$ can occur only for $\lambda\le0$ and implies $S_2=0$ (recall \eqref{kappa}).

\subsubsection{$\lambda<0$: (generalized) Siklos waves}

\label{subsubsec_Siklos}

Here $\lambda\equiv-1/l^2<0$ and $\a^2=l^2\beta_\sigma\beta^\sigma\neq0$. With \eqref{u_redef} one can set $\alpha=1$, so that here the canonical form is
\be
 \kappa=0 , \qquad l^{-2}=\beta_\sigma\beta^\sigma , \qquad Q=\frac{1}{4l^2}(x_\sigma+2l^2\b_\sigma)(x^\sigma+2l^2\b^\sigma) .
 \label{Q3}
\ee

This now enables one to integrate~\eqref{typeN_-1_final} and show \cite{Voldrich_master} that $Z_{[\a,\b]}$ is again of the form~\eqref{dZ_pp} (but now with $Q$ given by~\eqref{Q3}). By the same argument as in section~\ref{subsec_genpp}, one can then transform away $Z_\a$.

An important special subcase (not permitted in families with $\kappa\neq0$) arises when $\bl=\pa_v$ is a Killing vector field, i.e., when $L=F^{-1}S_1=0$, cf.~\eqref{Lg} (see also section~\ref{subsec_summary}). For the latter geometries, the alternative coordinates described in appendix~\ref{app_Siklos} reveal that these are higher-dimensional extensions of the well-known four-dimensional metrics of \cite{Siklos85} (see also \cite{OzsRobRoz85,Podolsky98sik,BicPod99I}), and we will thus refer to them as ``Siklos waves''. The general metrics with $L\neq0$ will be instead dubbed ``generalized Siklos waves'' (as in the four-dimensional case \cite{GriDocPod04}).

\subsubsection{$\lambda=0$: \pp waves}

\label{subsubsec_pp}

Here necessarily $\beta_\sigma=0\neq\alpha$, so that $Q=\alpha(u)$ and $P=1$. By \eqref{u_redef} one can set $Q=1$. From \eqref{typeN_-1_S1} one has $S_{1|i}=0$, so that by a transformation $u\mapsto f(u)$, $v\mapsto v/f'(u)$ one can achieve $S_{1}=0$ \cite{Coleyetal06}, such that $S_{,v}=0$ and $\bl=\pa_v$ is a covariantly constant vector field. Furthermore, eq.~\eqref{typeN_-1_final} gives $Z_{[j||k]i}=0$, which ensures that a coordinate transformation exists by which one can set $Z_i=0$ \cite{Schimming74} (cf. also \cite{Horowitz90,Sokolowski91,KucOrt19,OrtPraPra24}). One is thus left with the canonical form of \pp waves of type N 
\be
 \d s^2=2\d u\left[S_0(u,x)\d u+\d v\right]+\delta_{\a\b}\d x^\a\d x^\b ,
 \label{pp}
\ee
which belong to the more general class of spacetimes of \cite{Brinkmann25}. In the special case of conformally flat spacetimes (i.e., of Weyl type O), the function $S_0$ can be reduced to the form \cite{KucOrt19} 
\be
 S_0=s(u)x_\a x^\a .
\ee 
See also \cite{Schimming74,Coleyetal06} for further discussions.

\subsection{Summary: final form of the metric and relation to the (A)dS-Kerr-Schild class}

\label{subsec_summary}

To summarize, we have shown above in sections~\ref{subsec_k>0}, \ref{subsec_genpp} and \ref{subsec_Siklos} that for the three invariant subfamilies of Kundt metrics of Weyl and traceless-Ricci type N one can always locally set (in the coordinates~\eqref{metric})
\be
 Z_\a=0 . 
 \label{Z=0}
\ee
The remaining metric functions are specified by \eqref{S}, \eqref{g_canon}, \eqref{F}, \eqref{kappa}, while $S_0(u,x)$ remains arbitrary. In turn, the canonical form of $Q$ appearing in these equations is defined in table~\ref{tab_canonical} in the three possible cases, while integration of~\eqref{typeN_-1_S1} gives  
\be
 S_1=\frac{Q_{,u}}{Q} .
 \label{S1}
\ee
This means, in particular, that $S_1$ can be non-zero only in the case $\kappa=0$, i.e., for (generalized) Siklos waves. 
(The coordinate freedom \eqref{u_redef}, \eqref{v_redef} with $f=0$, i.e., $u=u(u')$, $v=h(u')v'$ has been used to set to zero in~\eqref{S1} an arbitrary additive function of $u$, taking $\d u/\d u'=h^{-1}$ in order to preserve the canonical form of the line-element).

By construction, the only non-zero components of the corresponding Riemann tensor are given by~\eqref{typeN_0} and (from~\eqref{Riem-2} with \eqref{Z=0}, \eqref{n_m}, \eqref{n})
\be
	R_{1i1j}=-F^{-1}\bigg[F^{-1}F_{|(i}S_{0|j)}+S_{0||ij}\bigg] ,   
	\label{R1i1j}
\ee
where we have got rid of the term $S_1$ of eq.~\eqref{S1} thanks to footnote~\ref{foot_Q}. This reveals that the $S_0=0$ limit of the line-element, i.e., 
\be
 \d s_0^2=2\frac{Q^2}{P^2}\d u\left[\left(\frac{\kappa}{2}v^2+\frac{Q_{,u}}{Q}v\right)\d u+\d v\right]+P^{-2}\delta_{\a\b}\d x^\a\d x^\b ,
 \label{background}
\ee
describes a spacetime of constant curvature, with Ricci scalar given by~\eqref{lambda}. The final form of the full metric is thus
\be
 \d s^2=\d s_0^2+2\frac{Q^2}{P^2}S_0(u,x)\d u^2 ,
 \label{KS_form}
\ee 
which demonstrates it belongs to the (A)dS-Kerr-Schild class.

Combined with Proposition~4 of \cite{MalPra11}, this proves that {\em Kundt spacetimes of Weyl and traceless-Ricci type N coincide with the class of non-expanding (A)dS-Kerr-Schild metrics}.\footnote{In passing, this conclusion reveals that the class of metrics introduced in \cite{GurSisTek12,GurSisTek14} (defined there as a particular subset of the Kerr-Schild Kundt metrics, cf. also \cite{Gulluetal11,GurSisTek17}) is in fact {\em equivalent} to the Kundt spacetimes of Weyl and traceless-Ricci type N. The inclusion of the former in the latter was noticed already in \cite{GurSisTek14,GurSisTek17}.\label{footn_GurSisTek12}} This was already known in the special case $\lambda=0$ \cite{OrtPraPra09}, corresponding to (non-expanding) Kerr-Schild spacetimes,\footnote{More precisely, reference \cite{OrtPraPra09} focused on Ricci flat spacetimes. However, the proof presented there covers also the Ricci type~N case.} and for any $\lambda$ when $D=4$ \cite{MalPra11} (see also \cite{Ortaggio18cqg}).  
One can further note that $\pa_v$ is a Killing field iff $\kappa=0$ and $Q_{,u}=0$, which thus defines the special subset of Siklos waves (for which $\beta^\sigma_{\phantom{\sigma},u}=0$; cf. also appendix~\ref{app_Siklos}).

For later use, let us write down explicitly also the Ricci tensor, which reads
\be
 R_{ab}=(n+1)\lambda g_{ab}+R_{11}\ell_a\ell_b , \qquad R_{11}=-\frac{P^2}{Q^2}\Box S_0 ,
 \label{Ricci_final}
\ee
where $\Box$ is the D'Alembert operator of the $D$-dimensional metric~\eqref{KS_form}. When $\Box$ acts on a $v$-independent function $f$, one has
\be
 \Box f=P^2\left[\pa_{\alpha}\pa_{\alpha}+\pa_{\alpha}\left(\ln\frac{Q^2}{P^{n}}\right)\pa_{\alpha}\right]f \qquad (f_{,v}=0) ,
 \label{Box}
\ee
such that, in that case, $\Box$ effectively reduces to the D'Alembert operator of the constant curvature background $\d s_0^2$.

As noticed in \cite{Obukhov04}, for certain applications it is useful to reexpress $R_{11}$ in terms of a rescaled function
\be
 H\equiv-\frac{2Q}{P^{n/2}}S_0 ,
\ee
such that\footnote{To arrive at this compact form of $R_{11}$, it is convenient to express the second covariant derivative of $Q$ using the curvature condition $R_{0i1j}=\lambda\delta_{ij}$ (imposed in section~\ref{sec_N}, cf. \eqref{typeN_0}) with \eqref{Riem0}, \eqref{F}. Equivalent expressions can be found in \cite{Obukhov04,GleDot05}.} 
\be
 R_{11}=\frac{P^{n/2+4}}{2Q^3}\left[H_{,\a\a}+\lambda\frac{n(n+2)}{4P^2}H\right] .
 \label{R11_H}
\ee
In terms of $H$, metric~\eqref{KS_form} takes the form~\eqref{metric_final}, \eqref{P}, as announced in section~\ref{sec_intro}.

\section{Applications in various gravity theories}

\label{sec_examples}

The results of the previous sections amount to a purely geometric (i.e., ``off shell'') classification of all $D$-dimensional Kundt metrics of Weyl and traceless-Ricci type N. In this section we will briefly discuss the role of such spacetimes as solutions of certain gravity theories of particular interest, naturally starting from general relativity in vacuum and with null electromagnetic fields.

\subsection{General relativity}

\subsubsection{Vacuum and universal spacetimes}

\label{subsubsec_vacGR}

In Einstein's theory, metric~\eqref{metric_final}, \eqref{P} in general represents a spacetime filled by (aligned) pure radiation, cf.~\eqref{Ricci_final}, \eqref{R11_H} (see also \cite{Obukhov04}). In vacuum, Einstein's equations reduce to 
\be
	H_{,\a\a}+\lambda\frac{n(n+2)}{4P^2}H=0 ,
	\label{vacuum_eq}
\ee	
where $\lambda$ is related to the Ricci scalar by~\eqref{lambda}.

Such spacetimes describe non-expanding gravitational waves with wave fronts (i.e., the $n$-dimensional surfaces of constant $u$ and $v$) of constant curvature, with spatial Ricci scalar $\tilde R=n(n-1)\lambda$. 
Some particular higher-dimensional solutions have been obtained in \cite{Obukhov04} and the general solution in \cite{GleDot05} (for $D=4$ see the earlier works \cite{OzsRobRoz85,Siklos85}). Solutions for the special case $\kappa=0$, $\lambda<0$ with $Q_{,u}=0$ (i.e. Siklos waves, cf. section~\ref{subsubsec_Siklos}) have been discussed earlier \cite{GibRub86,ChaGib00} using the parametrization of appendix~\ref{app_Siklos}. In particular, the higher-dimensional extension of the Kaigorodov metric \cite{Kaigorodov63} was given in \cite{CveLuPop99}.

It should be observed that for Kundt metrics of Weyl type N that additionally solve vacuum Einstein's gravity, any possible higher-order ``corrections'' to the field equations vanish identically, i.e., the metrics constructed above belong to the class of universal spacetimes \cite{HerPraPra14} (see also \cite{Deser75,Guven87,AmaKli89,HorSte90,Horowitz90,HorItz99,Coleyetal08,Gursesetal13} for earlier results in special cases). This means that they are also vacuum solutions of virtually any diffeomorphism invariant theory of gravity formulated in the context of Lorentzian geometry. 

However, by dropping the Einstein condition~\eqref{vacuum_eq}, solutions specific to some particular theories (i.e., non-universal spacetimes) can also be obtained. These are discussed in the following sections in a few cases of special interest.

\subsubsection{Maxwell $p$-forms}

\label{subsubsec_Maxw}

Beyond vacuum general relativity, the most natural matter content compatible with the assumptions of the present paper consists of null $p$-form fields aligned with the null direction of $\bl$ and satisfying linear Maxwell-like equations. In the geometries~\eqref{metric_final}, \eqref{P}, these can be written as \cite{OrtPra16}
\be
 \bF=\frac{1}{(p-1)!}f_{\alpha_1\ldots\alpha_{p-1}}(u,x)\d u\wedge\d x^{\alpha_1}\wedge\ldots\wedge\d x^{\alpha_{p-1}} ,
 \label{F_N_coords}
\ee
where $f_{\alpha_1\ldots\alpha_{p-1}}$ are the components of a $(p-1)$-form subject to the (reduced) Maxwell equations
\be
   f_{[\a_2\ldots\a_{p-1},\a_1]}=0 , \qquad  \left(P^{2(p-1)-n}f_{\beta\a_1\ldots\a_{p-2}}\right)_{,\beta}=0 ,
	\label{Maxwell}
\ee
while their $u$-dependence is arbitrary. Keeping into account the backreaction of $\bF$, Einstein's equations are equivalent to
\be
	H_{,\a\a}+\lambda\frac{n(n+2)}{4P^2}H=2\kappa_0 \frac{P^{2(p-1)-n/2}}{Q}(f_{\beta_1\ldots\beta_{p-1}}f_{\beta_1\ldots\beta_{p-1}}) ,
\ee	
where $\kappa_0$ is (up to a numerical factor) Newton's constant.

Solutions to the above equations represent a non-expanding gravitational wave accompanied by an electromagnetic wave. For the standard Maxwell case $p=2$, the first of~\eqref{Maxwell} means that, locally, $f_{\a}=f_{,\a}$, for some scalar function $f(u,x)$, and one can take $\bF=\d\bA$, with $\bA=-f\d u=-fF^{-1}\ell_a\d x^a$; this case  
has been discussed for $D=4$ in \cite{OzsRobRoz85} and in arbitrary dimension in \cite{Obukhov04} (cf. \cite{GleDot05} for a Yang-Mills extension), where more details on the solutions can be found. The special choice $2p=D$ (in even dimensions) corresponds to conformally invariant Maxwell equations (cf. the second of \eqref{Maxwell}).

\subsubsection{Non-linear electrodynamics}

Thanks to the fact that $\bF$ in~\eqref{F_N_coords} is null, the Einstein-Maxwell solutions of section~\ref{subsubsec_Maxw} also solve general relativity coupled to (virtually) any non-linear $p$-form electrodynamics \cite{KucOrt19}.\footnote{By which we mean that ``corrections'' to the corresponding action are constructed from algebraic invariants of $\bF$, i.e., without taking covariant derivatives -- cf. \cite{KucOrt19} for a more precise formulation. Beware possible pathologies of null fields in particular theories such as ModMax electrodynamics \cite{Bandosetal20}.} For the particular case $D=4=2p$, these results date back to \cite{Schroedinger35,Schroedinger43,Kichenassamy59,KreKic60,Peres61} and provide exact solutions, for example, to Einstein gravity coupled to the well-known Born-Infeld electrodynamics \cite{Born33,BorInf34}. 

For specific non-linear theories of electrodynamics, one can also find solutions that are {\em not} contained in those of section~\ref{subsubsec_Maxw}. Let us mention here only a case of particular interest, namely the conformally invariant 2-form electrodynamics put forward in \cite{HasMar07}, for which the matter part of the Lagrangian is proportional to $(F_{ab}F^{ab})^{D/4}$. For such a theory, null fields are stealth (i.e., they do not backreact) and satisfy the generalized Maxwell equations identically provided $\bF=\d\bA$ \cite{KokOrt21}. This means that any Einstein spacetime~\eqref{metric_final}, \eqref{P}, \eqref{vacuum_eq} with $\bF$ as in~\eqref{F_N_coords} (with $p=2$) is a solution provided the single condition $f_{[\a_2,\a_1]}=0$ is fulfilled.

\subsection{Gauss-Bonnet and Lovelock gravity}

\label{subsec_Lovelock}

Lovelock gravity defines the most general class of theories 
for which the field equations amount to the vanishing of a symmetric, divergence-free, rank-2 tensor constructed from the metric and its first two derivatives \cite{Lovelock71}. In vacuum they read 
\be
 \sum_{k=0}^{[(D-1)/2]}c_k G^{a(k)}_{c}=0, \qquad\qquad G^{a(k)}_{c}\equiv-\dfrac{1}{2^{k+1}}\delta^{aa_1b_1\ldots a_kb_k}_{cc_1d_1\ldots c_kd_k}R^{c_1d_1}_{a_1b_1}\ldots R^{c_kd_k}_{a_kb_k} ,
 \label{fieldeqns}
\ee
where $\delta^{a_1\ldots a_p}_{c_1\ldots c_p}=p!\delta^{a_1}_{[c_1}\ldots\delta^{a_p}_{c_p]}$, and $c_k$ are coupling constants. Special choices of the latter give rise, e.g., to Einstein's ($c_k=0$ for $k>1$) or Gauss-Bonnet's theory ($c_k=0$ for $k>2$), while the cosmological constant is $-c_0$ (up to normalization). For $D=3,4$ only the terms $k=0,1$ are non-vanishing in~\eqref{fieldeqns}.

Since for the class of geometries~\eqref{metric_final}, \eqref{P} the Riemann tensor obeys~\eqref{typeN_0}, \eqref{typeN_-1}, the vacuum equations~\eqref{fieldeqns} simplify drastically and reduce to \cite{GleDot05} (see also the appendix of~\cite{Ortaggio18prd})
 \beq 
  & & p(\lambda)=0 , \qquad  (D-3)(D-4)p'(\lambda)R_{11}=0 , \label{eqs_Lovel} \\
  & & p(\lambda)\equiv \sum_{k=0}^{[(D-1)/2]} c_k\frac{\lambda^k}{(D-2k-1)!} ,
\eeq 
where $R_{11}$ is given by~\eqref{R11_H} and $p'\equiv\d p/\d\lambda$. The first of~\eqref{eqs_Lovel} determines $\lambda$ in terms of the $c_k$, exactly as in the case of constant curvature spacetimes \cite{BouDes85,Wheeler86}. The second of~\eqref{eqs_Lovel} means that generically $R_{11}=0$, i.e., the spacetime must be Einstein \cite{GleDot05}, giving rise to the same solutions as in general relativity (section~\ref{subsubsec_vacGR}). On the other hand, for {\em degenerate} Lovelock theories where $\lambda$ is taken to be (at least) a double root of $p(\lambda)$, the second of~\eqref{eqs_Lovel} is satisfied identically and the metric function $H$ remains arbitrary \cite{GleDot05}. In the special case of Siklos waves (i.e., metrics~\eqref{metric_final}, \eqref{P} with $\kappa=0$, $\lambda<0$, $Q_{,u}=0$), it was already recognized in \cite{GibRub86} that Einstein spacetimes also solve the Einstein-Gauss-Bonnet theory.

See \cite{GleDot05} for the possible inclusion of electromagnetic and Yang-Mills fields, and \cite{Ortaggio18prd} for comments on vacua of Weyl and traceless-Ricci type~III.

\subsection{Quadratic gravity}

\label{subsec_QG}

The Einstein-Hilbert action augmented by the most general term quadratic in the curvature gives rise to the field equations \cite{Gregory47,Buchdahl48} (in the notation of \cite{DesTek03}) 
\beq
  \frac{1}{\kappa_0}\left( R_{a b} - \frac{1}{2} R g_{a b} + \Lambda g_{a b} \right)
  + 2 \alpha R \left( R_{a b} - \frac{1}{4} R g_{a b} \right)
  + \left( 2 \alpha + \beta \right)\left( g_{a b} \Box - \nabla_a  \nabla_b \right) R \nonumber \\
	{}+ \beta\left[\Box\left( R_{a b} - \frac{1}{2} R g_{a b} \right)
  + \left(2R_{a c  b d} - \frac{1}{2} g_{a b} R_{c  d} \right) R^{c  d}\right]  \nonumber \\
  {}+ 2 \gamma \left[ R R_{a b} - 2 R_{a c  b d} R^{c  d}
  + R_{a c  de} R_{ b}^{\phantom{ b}c  de} - 2 R_{a c } R_{ b}^{\phantom{ b}c }
  - \frac{1}{4} g_{a  b}\left(R_{cdef}R^{cdef}-4R_{cd}R^{cd}+R^2 \right)\right]=0 ,
  \label{E_QG}
\eeq
where $\Lambda$ is the cosmological constant and $\alpha$, $\beta$, $\gamma$ are parameters specifying how quadratic terms enter the action. In particular, $\gamma$ corresponds to the Gauss-Bonnet term (which also entered~\eqref{fieldeqns} as $G^{a(2)}_{c}$).

For the metrics~\eqref{metric_final}, \eqref{P}, using~\eqref{typeN_0}, \eqref{typeN_-1}, \eqref{lambda}, \eqref{R1i1j} and \eqref{Ricci_final}, the vacuum equations~\eqref{E_QG} reduce to \cite{MalPra11prd}\footnote{The form of the field equation~\eqref{eqs_QG2} is given in \cite{MalPra11prd} in the Newman-Penrose formalism, without using the explicit metric~\eqref{metric_final}, \eqref{P}.}
 \beq 
  & & \frac{1}{\kappa_0} \left[(D-2)\lambda-\frac{2\Lambda}{D-1}\right]+\lambda^2(D-4)\big[(D-1)(D\alpha+\beta)+\gamma (D-2)(D-3)\big]=0 , \label{eqs_QGl} \\
  & & \left\{\frac{1}{\kappa_0}+2\lambda\big[D(D-1)\alpha+(D-3)(D-4)\gamma\big]+\beta\big[\Box+2P^2(\ln F)_{,\alpha}\pa_\alpha+6F^{-1}\kappa-4\lambda\big]\right\}R_{11}=0 , \label{eqs_QG2}
\eeq 
with~\eqref{P}, \eqref{F}, \eqref{Q}, \eqref{kappa}, \eqref{Ricci_final}, \eqref{Box}.

Eq.~\eqref{eqs_QGl} determines $\lambda$ in terms of the parameters of the theory (there being generically two branches), while \eqref{eqs_QG2} is a fourth-order equation for the metric function $S_0$. As noticed in \cite{MalPra11prd}, Einstein spacetimes clearly satisfy~\eqref{eqs_QG2} identically, while in the special case $\beta=0$, eq.~\eqref{eqs_QG2} becomes an algebraic constraint on the parameters and $S_0$ remains arbitrary. The particular case of Siklos waves (i.e., $\kappa=0$, $\lambda<0$, $Q_{,u}=0$) has been discussed in more detail in \cite{AliFar11,Gulluetal11,MalPra11prd} in the coordinates of appendix~\ref{app_Siklos}. Generalized \pp waves (i.e., $\kappa=\lambda$) have been studied in \cite{GurSisTek12}
 in the case $\lambda<0$ using the coordinates of appendix~\ref{app_gen_pp}.

\subsection{$f(R)$ gravity}

\label{subsec_f(R)}

A relatively simple and widely explored theory which goes beyond second order powers of the curvature is given by the so called $f(R)$ gravity, originally discussed from a cosmological viewpoint \cite{Buchdahl70}. In the metric approach, it is described by the following equations of motion \cite{Buchdahl70}
\beq
  \left(R_{\mu\nu}+g_{\mu\nu}\Box-\nabla_\mu\nabla_\nu\right)f'-\frac{1}{2}fg_{\mu\nu}=0 ,
	\label{f(R)}
\eeq
where $f$ is an arbitrary function of the Ricci scalar $R$ and $f'=\pa f/\pa R$.

Since the geometries~\eqref{metric_final}, \eqref{P} have $R=$const (cf.~\eqref{lambda}), eq.~\eqref{f(R)} generically implies that the spacetime is Einstein (with $H$ determined by~\eqref{vacuum_eq}), and the constant $R$ must satisfy the condition (cf. \cite{BarOtt83} for related comments when $D=4$)
\be
  2Rf'=Df .
	\label{f(R)_Einstein}
\ee
However, for theories such that $f(R)$ admits a double zero at $R=\bar R=$const, (i.e., $f(\bar R)=0=f'(\bar R)$, cf. \cite{NojOdi14} for related comments in a different context), eq.~\eqref{f(R)} is satisfied identically by any metric having $R=\bar R$, hence non-Einstein spacetimes are also solutions, and $H$ remains undetermined in~\eqref{metric_final}.

For example, for polynomial theories $f(R)=\sum_{k=0}c_kR^k$, eq.~\eqref{f(R)_Einstein} becomes
\be
 \sum_{k=0}\left(D-2k\right)c_kR^k=0 ,
\ee
determining $R$ (i.e., $\lambda$) in terms of the constants $c_k$ ($R$ remains arbitrary in the scale invariant case where $c_{D/2}$ is the only non-zero coefficient in $f(R)$ \cite{Buchdahl48_3}.) On the other hand, the non-Einstein solutions mentioned above require the existence of a constant $\bar R$ such that
\be
 \sum_{k=0}c_k\bar R^k=0=\sum_{k=0}kc_k\bar R^{k-1} .
 \label{tuning_F}
\ee

\subsection{More general higher-order gravities and almost universal spacetimes}

The reader may have noticed that, for all the generalized gravity theories considered in sections~\ref{subsec_Lovelock}, \ref{subsec_QG} and \ref{subsec_f(R)}, the vacuum field equations for metrics~\eqref{metric_final}, \eqref{P} always reduce to an algebraic equation relating $\lambda$ to the parameters of the theory under consideration, accompanied by a linear PDE for the metric function $H$ (except for a few ``degenerate'' cases where $H$ remains arbitrary). It should be remarked that this is no coincidence, as follows from \cite{Gursesetal13,Kuchynkaetal19}. In fact, all line-elements~\eqref{metric_final}, \eqref{P} belong to the class of ``almost universal'' spacetimes \cite{Kuchynkaetal19}. The results of \cite{Gursesetal13,Kuchynkaetal19} thus imply that a similar form of the vacuum field equations would actually hold for any gravity theory defined by a Lagrangian constructed from the metric, the Riemann tensor and its covariant derivatives of arbitrary order. Further discussion and examples can be found in \cite{Gursesetal13,Kuchynkaetal19}.

Going beyond vacuum spacetimes, one may mention, e.g., the study of metrics~\eqref{metric_final}, \eqref{P} in \cite{GurHeySen21} (in a different notation, cf. footnote~\ref{footn_GurSisTek12}) as solutions of gravity coupled non-minimally to a 2-form via terms of the form proposed in \cite{Prasanna71} or in \cite{Horndeski76}, as well as the earlier results \cite{Horndeski79,GurHal78} for $D=4$ and $\lambda=0$.

\section*{Acknowledgments}

We are grateful to Vojt\v ech Pravda and Alena Pravdov\' a for reading the manuscript. M.O. has been supported by the Institute of Mathematics, Czech Academy of Sciences (RVO 67985840). The work of J.B. is supported by FONDECYT Postdoctorado grant 3230596.

\renewcommand{\thesection}{\Alph{section}}
\setcounter{section}{0}

\renewcommand{\theequation}{{\thesection}\arabic{equation}}
\setcounter{equation}{0}

\section{Conservation equation in Kundt spacetimes}

\label{app_conserv}

In this appendix we discuss certain properties of the structure of the equations of motions for general Kundt spacetimes (i.e., without assuming any extra conditions on their curvature, as opposed to what was done in sections~\ref{sec_kundt} and \ref{sec_N}) in an arbitrary diffeomorphism-invariant theory of gravity.\footnote{A corresponding analysis for Robinson-Trautman spacetimes in arbitrary dimension can be found in appendix~A of \cite{KokOrt21}.} By doing so, we clarify and generalize certain results obtained previously in special cases. We proceed in a coordinate-independent way. 

Let us set up a null frame $\{\bE_0,\bE_1,\bE_i\}$ such that $\bE_0=\bl$ is aligned with the Kundt vector field. From the definition~\eqref{kundt} of Kundt spacetimes with the affine parameter condition~\eqref{affine} one readily obtains (cf. also, e.g., appendix~A of \cite{OrtPra16} and references therein)
\be
 \Gamma^1_{\phantom{1}i0}=0 , \qquad \Gamma^1_{\phantom{1}ij}=0 , \qquad \Gamma^1_{\phantom{1}10}=0 .
 \label{kundt2}
\ee

Without losing generality \cite{OrtPraPra07}, one can always rotate the vectors $\bE_i$ such that, in addition,\footnote{Both \eqref{kundt2} and \eqref{kundt2_Gammaij0} are satisfied, in particular, for the special sub-family studied in section~\ref{sec_N}, cf. \eqref{G11}--\eqref{Gij}.}
\be
 \Gamma^i_{\phantom{i}j0}=0 .
 \label{kundt2_Gammaij0}
\ee

Let us consider a diffeomorphism-invariant theory of gravity of the form 
\be
 S=\int\d^dx\sqrt{-g}{\cal L}(\bR,\nabla\bR,\ldots) ,
\label{action}
\ee
where ${\cal L}$ is a scalar invariant constructed polynomially from the Riemann tensor $\bR$ and its covariant derivatives of arbitrary order. 

The corresponding equations of motion read $\ensuremath{\boldsymbol{E}}=0$, where $\ensuremath{\boldsymbol{E}}$ is a symmetric 2-tensor defined by 
\cite{Eddington_book}
\be
 E_{ab}\equiv\frac{1}{\sqrt{-g}}\frac{\delta\left(\sqrt{-g}\cal L\right)}{\delta g^{ab}} . \label{grav_eq} 
\ee

Furthermore, $\ensuremath{\boldsymbol{E}}$ is divergenceless, i.e., $E^{ab}_{\phantom{ab};b}=0$  \cite{Eddington_book}. This property enables one to remove some redundancy in the field equations for a Kundt spacetime in any theory of the form~\eqref{action}. Namely, using \eqref{kundt2} and \eqref{kundt2_Gammaij0} the condition $E^{ab}_{\phantom{ab};b}=0$ can be written for $a=1,i,0$ as
\beq 
 & & E^{1b}_{\phantom{1b}|b}+E^{11}(\Gamma^1_{\phantom{1}11}+\Gamma^b_{\phantom{b}1b})+E^{1i}(\Gamma^1_{\phantom{1}1i}+\Gamma^1_{\phantom{1}i1}+\Gamma^b_{\phantom{b}ib})=0 , \label{bianchi+1} \\
 & & E^{ib}_{\phantom{ib}|b}+E^{11}\Gamma^i_{\phantom{i}11}+E^{1j}(\Gamma^i_{\phantom{i}1j}+\Gamma^i_{\phantom{i}j1})+E^{10}(\Gamma^i_{\phantom{i}01}+\Gamma^i_{\phantom{i}10})+E^{1i}\Gamma^b_{\phantom{b}1b}+E^{jk}\Gamma^i_{\phantom{i}jk}+E^{ij}\Gamma^b_{\phantom{b}jb}=0 , \label{bianchi0} \\
 & & E^{0b}_{\phantom{1b}|b}+E^{1i}\Gamma^0_{\phantom{0}i1}+E^{01}(\Gamma^0_{\phantom{0}01}+\Gamma^b_{\phantom{b}1b})+E^{ij}\Gamma^0_{\phantom{0}ij}+E^{0i}(\Gamma^0_{\phantom{0}0i}+\Gamma^0_{\phantom{0}i0}+\Gamma^b_{\phantom{b}ib})=0 . \label{bianchi-1}
\eeq

In Kundt spacetimes it is usually most convenient to solve the field equations in order of decreasing ``boost weight'' (b.w.) \cite{Milsonetal05,OrtPraPra13rev}, i.e., in the order $E^{11}=0$ (b.w.~$+2$), $E^{1i}=0$ (b.w.~$+1$), $E^{10}=0=E^{ij}$ (b.w.~0), $E^{0i}=0$ (b.w.~$-1$) and finally $E^{00}=0$ (b.w.~$-2$). Assuming one has already solved the equations of positive b.w., eq.~\eqref{bianchi+1} then reduces to the simple identity
\be
 E^{10}_{\phantom{10}|0}=0 . 
 \label{id1}
\ee

Similarly, once also the equations of zero b.w. have been solved, eq.~\eqref{bianchi0} becomes
\be
 E^{i0}_{\phantom{i0}|0}=0 . 
 \label{idi}
\ee

Finally, after the equations of b.w.~$-1$ have also been solved, eq.~\eqref{bianchi-1} gives the last identity
\be
 E^{00}_{\phantom{00}|0}=0 . 
 \label{id0}
\ee

In adapted coordinates such that $\bE_0=\pa_v$ (as used, e.g., in sections~\ref{sec_kundt} and \ref{sec_N}), the above identities~\eqref{id1}--\eqref{id0} mean that if any of the field equations $E^{10}=0$, $E^{i0}=0$ or $E^{00}=0$ consists of addends with a different $v$-dependence, all such addends vanish identically (and separately) as a consequence of the general conservation law $E^{ab}_{\phantom{ab};b}=0$ (provided in each case the field equations of higher b.w. have been already solved, as explained above), except for the possible $v$-independent addends. This observation enables one to immediately get rid of a redundancy of the field equations in a theory-independent way, with no need of otherwise laborious computations on a case-by-case basis (as done for special cases in Einstein's gravity, for example, in footnote~8 of \cite{OrtPra16} or in \cite{Krtousetal12} -- see also chapter~31 of \cite{Stephanibook} in four dimensions).

The above discussion extends essentially unchanged to actions containing also matter fields (including non-minimally coupled ones) -- indeed the conservation law $E^{ab}_{\phantom{ab};b}=0$ still holds, provided the equations of motions for the matter fields are satisfied \cite{Eddington_book,Horndeski74_unp,Horndeski76,Anderson78,IyeWal94}.

\section{Kundt coordinates for all Kundt spacetimes of Weyl and traceless-Ricci type N}

\label{app_kundt_coords}

The substitution  \cite{BicPod99I} (cf. also, e.g., \cite{PodOrt03,GriPodbook})
\be
 v=F^{-1}r 
\ee
brings the line-element~\eqref{KS_form} with \eqref{background} of Kundt spacetimes of Weyl and traceless-Ricci type N into the standard Kundt coordinates \cite{Coleyetal03,ColHerPel06,PodOrt06,PodZof09,OrtPraPra13rev,PodSva13}
\be
 \d s^2=2\d u(\d r+{\cal H}\d u+W_\g\d x^\g)+P^{-2}\delta_{\a\b}\d x^\a\d x^\b \qquad \left(F=\frac{Q^2}{P^2}\right) ,
\ee 
with $P$ given by~\eqref{P} and 
\be
 {\cal H}=\frac{\kappa}{2}F^{-1}r^2-\frac{Q_{,u}}{Q}r+FS_0(u,x) , \qquad W_\g=-r(\ln F)_{,\g} .
\ee 

Using~\eqref{Q}, one can write explicitly
\be
 (\ln F)_{,\g}=\frac{2}{Q}\b_\g-\frac{\lambda}{PQ}x_\g(2\a+\b_\sigma x^\sigma) .
\ee  
The respective canonical forms of $\kappa$ and $Q$ (and thus of $\a$ and $\b_\sigma$) for the three invariantly defined subfamilies of solutions are given in table~\ref{tab_canonical}.

\section{Alternative coordinates for Siklos waves ($\kappa=0$, $\lambda<0$, $Q_{,u}=0$)}

\label{app_Siklos}

For the subclass of geometries given by the third row of table~\ref{tab_canonical} with the additional condition $\dot\beta^\sigma=0$ (i.e., $Q_{,u}=0$, which by~\eqref{S1} gives $S_1=0$ and thus makes $\pa_v$ a Killing vector field, cf. section~\ref{subsubsec_Siklos}), it is useful to consider also an alternative system of coordinates, which is often used in the literature and dates back to \cite{Siklos85} in four dimensions.

First, since $\beta^\sigma$ does not depend on $u$, we can use a $u$-independent rotation~\eqref{rotation} to set in~\eqref{Q3}
\be
 \b_\sigma=l^{-1}\delta_{\sigma,1} .
 \label{beta_Sikl}
\ee
Then, the coordinate transformation
\be
 x_1=2l\frac{l^2-y_\rho y^\rho}{2ay_1+a^2+y_\sigma y^\sigma} , \qquad x_{\hat\a}=y_{\hat\a}\frac{4l^2}{2ay_1+a^2+y_\sigma y^\sigma} \quad (\hat\a\neq1) , 
 \label{transf_Sikl}
\ee
brings the line-element~\eqref{KS_form} with \eqref{background}, \eqref{g_canon} \eqref{Q3}, \eqref{beta_Sikl} with \eqref{Q3}, \eqref{beta_Sikl} into the form
\be
	\d s^2=\frac{l^2}{y_1^2}\left(2\d u\d v+\delta_{\a\b}\d y^\a\d y^\b+2S_0\d u^2\right) , 
	\label{Siklos}
\ee	
where $\a,\b=1,\ldots,n$ and $S_0$ is an arbitrary function of $(u,y_1,\ldots,y_n)$.

Arbitrary dimensional metrics of the form~\eqref{Siklos} were first considered in \cite{GibRub86}, cf. also \cite{CveLuPop99,ChaGib00}. These spacetimes are Einstein when (recall~\eqref{Ricci_final}) \cite{GibRub86,ChaGib00}
\be
  y_1^n(y_1^{-n}S_{0,y_1})_{,y_1}+S_{0,y_{\hat\a}y_{\hat\a}}=0 .
\ee
In particular, the extension of the homogenous Kaigorodov solution \cite{Kaigorodov63} (see also \cite{Siklos85,Podolsky98sik}) corresponds to $S_0\sim y_1^{D-1}$ \cite{CveLuPop99,ChaGib00}.

In the special case $D=4$, the transformation~\eqref{transf_Sikl} becomes equivalent to the one given in \cite{Podolsky98sik}. For any $D$, the transformation of Siklos waves~\eqref{Siklos} to the Kundt coordinates of appendix~\ref{app_kundt_coords} can be found in \cite{ColFusHer09}.

\section{Alternative coordinates for generalized \pp waves ($\kappa=\lambda$) with $\lambda<0$}

\label{app_gen_pp}

The purpose of this appendix is to identify certain ``spherical wave'' metrics considered (for $\lambda<0$) in \cite{GurSisTek12} among the class of geometries studied in the present paper. Namely, we will display a coordinate transformation which brings the former into the generalized \pp waves of table~\ref{tab_canonical}. 

Since it is trivial to match the Kerr-Schild parts of our metrics (section~\ref{subsec_summary}) with those of \cite{GurSisTek12}, it suffices here to discuss only the background AdS metric $\d s_0^2$. The form of AdS used in \cite{GurSisTek12} reads
\be
 \d s_0^2=\frac{1}{|\lambda|r^2\cos^2\theta}(-\d t^2+\d r^2+r^2\d\Omega_n^2) ,
\ee
where $\d\Omega_n^2=\d\theta^2+\sin^2\theta\d\Omega_{n-1}^2$ is the standard metric of an $n$-dimensional round unit sphere (with $\d\Omega_{n-1}^2$ an $(n-1)$-dimensional round unit sphere). By the transformation
\be
 t=u-\frac{1}{\lambda v} , \qquad r=\frac{1}{\lambda v} , \qquad \cos\theta=\frac{1+\frac{\lambda}{4}\rho^2}{1-\frac{\lambda}{4}\rho^2} ,
\ee
one obtains
\be
 \d s_0^2=2\frac{\left(1-\frac{\lambda}{4}\rho^2\right)^2}{\left(1+\frac{\lambda}{4}\rho^2\right)^2}\d u\left(\frac{\lambda}{2}v^2\d u+\d v\right)+\frac{\d\rho^2+\rho^2\d\Omega_{n-1}^2}{\left(1+\frac{\lambda}{4}\rho^2\right)^2} .
 \label{gen_pp_rho}
\ee

Upon identifying $\rho=(x_\a x^\a)^{1/2}$ as a spatial radial coordinate in $n$ dimensions, it is clear that~\eqref{gen_pp_rho} is nothing but~\eqref{background} with 
\be
 P=1+\frac{\lambda}{4}x_\sigma x^\sigma , \qquad Q=1-\frac{\lambda}{4}x_\sigma x^\sigma , \qquad \kappa=\lambda ,
\ee
meaning that the corresponding full metric~\eqref{KS_form} is a generalized \pp wave. Note that, although the starting metric of \cite{GurSisTek12} assumed $\lambda<0$, the transformed generalized \pp wave admits any sign of $\lambda$ (including the \pp wave limit $\lambda=0$). The curvature of the wave fronts has the same sign as $\lambda$ (recall~\eqref{g_canon}).


\providecommand{\href}[2]{#2}\begingroup\raggedright\endgroup

\end{document}